\documentclass[aps,preprint,showpacs,nofootinbib]{revtex4}
\usepackage{graphicx}
\newcommand{\nc}{\newcommand}
\nc{\beq}{\begin{equation}}  \nc{\eeq}{\end{equation}}
\nc{\bea}{\begin{eqnarray}}  \nc{\eea}{\end{eqnarray}}
\nc{\baa}{\begin{array}}     \nc{\eaa}{\end{array}}
\nc{\bit}{\begin{itemize}}   \nc{\eit}{\end{itemize}}
\nc{\ben}{\begin{enumerate}} \nc{\een}{\end{enumerate}}
\nc{\bce}{\begin{center}}    \nc{\ece}{\end{center}}
\nc{\bpm}{\begin{pmatrix}}   \nc{\epm}{\end{pmatrix}}
\nc{\bvt}{\begin{verbatim}}  \nc{\evt}{\end{verbatim}}
\def\to{\rightarrow}

\def\gesim{\,{\raise-3pt\hbox{$\sim$}}\!\!\!\!\!{\raise2pt\hbox{$>$}}\,}
\def\lesim{\,{\raise-3pt\hbox{$\sim$}}\!\!\!\!\!{\raise2pt\hbox{$<$}}\,}
\def\boldoverdot{\,{\raise6pt\hbox{\bf.}\!\!\!\!\>}}

\def\lcal{{\cal L}}

\usepackage{bm}
\def\vev{vacuum expectation value}

\def\diag{\hbox{\diag}}

\def\gev{\hbox{GeV}}
\def\tev{\hbox{TeV}}

\def\vevof#1{\left\langle#1\right\rangle}
\def\doubleundertext#1{
{\undertext{\vphantom{y}#1}}\par\nobreak\vskip-\the\baselineskip\vskip4pt%
\undertext{\hbox to 2in{}}}
\def\inbox#1{\vbox{\hrule\hbox{\vrule\kern5pt
     \vbox{\kern5pt#1\kern5pt}\kern5pt\vrule}\hrule}}
\def\sqr#1#2{{\vcenter{\hrule height.#2pt
      \hbox{\vrule width.#2pt height#1pt \kern#1pt
         \vrule width.#2pt}
      \hrule height.#2pt}}}
\def\square{\mathchoice\sqr56\sqr56\sqr{2.1}3\sqr{1.5}3}
\def\today{\ifcase\month\or
  January\or February\or March\or April\or May\or June\or
  July\or August\or September\or October\or November\or December\fi
  \space\number\day, \number\year}
\def\pmb#1{\setbox0=\hbox{#1}%
  \kern-.025em\copy0\kern-\wd0
  \kern.05em\copy0\kern-\wd0
  \kern-.025em\raise.0433em\box0 }
\def\up#1{^{\left( #1 \right) }}
\def\inv#1{{1\over#1}}

\def\ui{U(1)}
\def\sumprime_#1{\setbox0=\hbox{$\scriptstyle{#1}$}
  \setbox2=\hbox{$\displaystyle{\sum}$}
  \setbox4=\hbox{${}'\mathsurround=0pt$}
  \dimen0=.5\wd0 \advance\dimen0 by-.5\wd2
  \ifdim\dimen0>0pt
  \ifdim\dimen0>\wd4 \kern\wd4 \else\kern\dimen0\fi\fi
\mathop{{\sum}'}_{\kern-\wd4 #1}}
\newcounter{problem}[section]

\nc{\non}{\nonumber}
\nc{\hc}{\hbox {h.c.}} 
\def\lsim{\mathrel{\raise.3ex\hbox{$<$\kern-.75em\lower1ex\hbox{$\sim$}}}}
\def\gsim{\mathrel{\raise.3ex\hbox{$>$\kern-.75em\lower1ex\hbox{$\sim$}}}}

\nc{\Lsp}{\;\;\;\;\;\;\;\;\;\;}  
\nc{\LLLsp}{\lspace \lspace}
\nc{\lsp}{\;\;\;\;\;\;}
\nc{\spac}{\;\;\;}
\nc{\noi}{\noindent}
\nc{\lra}{\longrightarrow}

\def\rad{L}      % radius of the compactified space
\def\cy{y}     % label for the compact coordiante
\def\qp{q_\psi}  % fermion charge
\def\e5{e_5}

\def\vevof#1{{\left\langle#1\right\rangle}}
\nc{\avev}{\vevof{A_4}}
\nc{\eps}{\epsilon}
\nc{\dcp}{\delta^{(CP)}_n}

\begin{document}
% ------------------------------------------------------------
   \def\thebibliography#1{\centerline{REFERENCES}
     \list{[\arabic{enumi}]}{\settowidth\labelwidth{[#1]}\leftmargin
     \labelwidth\advance\leftmargin\labelsep\usecounter{enumi}}
     \def\newblock{\hskip .11em plus .33em minus -.07em}\sloppy
     \clubpenalty4000\widowpenalty4000\sfcode`\.=1000\relax}\let
     \endthebibliography=\endlist
   \def\sec#1{\addtocounter{section}{1}\section*{\hspace*{-0.72cm}
     \normalsize\bf\arabic{section}.$\;$#1}\vspace*{-0.3cm}}
% ------------------------------------------------------------
\vspace*{-1.7cm}
\noindent
% {\large\bf October 1, 2003}
%
% \vspace{-0.7cm}
\begin{flushright}
$\vcenter{
\hbox{{\footnotesize CERN-PH-TH/2004-012}}
\hbox{{\footnotesize IFT-04-04}}
\hbox{{\footnotesize UCRHEP-T370}}
%\hbox{(hep-ph/????)}
%\hbox{\today}
}$
\end{flushright}

\vspace{1cm}
\bce
{\Large\bf CP Violation from 5-dimensional QED}

\vspace{1cm}

\renewcommand{\thefootnote}{\alph{footnote})}
{\sc Bohdan GRZADKOWSKI}\footnote{E-mail address:
\tt bohdan.grzadkowski@fuw.edu.pl}

\vspace*{0.1cm}
{\sl Institute of Theoretical Physics,  Warsaw University\\
Ho\.za 69, PL-00-681 Warsaw, Poland\\
and\\
CERN, Department of Physics\\
Theory Division\\
1211 Geneva 23, Switzerland}

\vspace*{0.5cm}
{\sc Jos\'e WUDKA}\footnote{E-mail address:
\tt jose.wudka@ucr.edu}

\vspace*{0.1cm}
{\sl Department of Physics, University of California\\
Riverside CA 92521-0413, USA}
\ece

\vspace*{0.4cm}
\centerline{ABSTRACT}

\vspace*{0.3cm}
\baselineskip=20pt plus 0.1pt minus 0.1pt

It has been shown that QED in $(1+4)$-dimensional space-time, with the fifth 
dimension compactified on a circle, leads to CP violation (CPV).
Depending on fermionic boundary conditions, CPV may be either explicit
(through the Scherk--Schwarz mechanism), 
or spontaneous (via the Hosotani mechanism). The fifth component of the 
gauge field acquires  (at the one-loop level) 
a non-zero vacuum expectation value. In the presence of two fermionic fields, this
leads to spontaneous CPV in the case of CP-symmetric boundary conditions.
Phenomenological consequences are illustrated by
a calculation of the electric dipole moment for the fermionic zero-modes.

\vspace*{0.4cm} \vfill



\newpage
%-------------------------------------------------------------
\renewcommand{\thefootnote}{$\sharp$\arabic{footnote}}
% \renewcommand{\thefootnote}{\sharp\arabic{footnote}}
%-------------------------------------------------------------
\pagestyle{plain} \setcounter{footnote}{0}
\baselineskip=21.0pt plus 0.2pt minus 0.1pt

% 111111111111111111111111111111111111111111111111111111111111
\section{Introduction}

The physics of grand unified theories has been plagued by 
fundamental difficulties to accommodate different mass scales within
a single theory, the so-called hierarchy problem.
For a long time supersymmetric models had the commendable
feature of being able to solve this problem. 
More recently,
non-supersymmetric higher-dimensional models were proposed~\cite{Randall:1999ee}, 
\cite{arkanietal}, which solve the hierarchy problem provided that
an appropriate space-time geometry is realized. 
Though in the original models only gravity was present outside
a 4-dimensional slice of the compactified space, this is not an inescapable restriction. In fact,
models where all fields propagate throughout the compactified
space-time are natural and phenomenologically viable~\cite{Antoniadis:1990ew}, \cite{ued}.
In this letter we consider quantum electrodynamics (QED) in 5 dimensions (5D)
focusing on the possibility that it naturally generates small but non-trivial
CP-violating effects\footnote{For earlier attempts to obtain CP violation within 
extra-dimensional extensions of the Standard Model (SM) of electroweak interactions, see 
Refs.~\cite{Chang:2001yn}, \cite{Cosme:2002zv}.}.

%22222222222222222222222222222222222222222222222222222222222222222222222222222222222222222222
\section{The Model}
We will consider  an
Abelian model in 5D, with coordinates 
$ x^M,~ M=0,\ldots,4$ and $ x^4=y$ compactified to a circle of radius $\rad$.
We assume the presence of two fermionic fields ($\psi_{1,2}$)
interacting with the U(1) gauge field $A_M$ according to the
Lagrangian
\beq
\lcal_{QED} =   -\inv4 F_{M N}^2
          + \sum_{i=1,2}\bar \psi_i\left( i \gamma^M D_M - M_i \right)\psi_i
          + \lcal_{gf}\, ,
\label{lagqed}
\eeq
where $ F_{M N} = \partial_M A_N - \partial_N A_M $,
the covariant derivative is given by
$ D_M = \partial_M + i e_5 q_i A_M $, where $q_i$ denotes the
charge of $\psi_i$ in units of $e_5$, and $\lcal_{gf}$ stands for a gauge-fixing term.
We will assume that the gauge fields
are periodic in $y$, but we will allow the fermions to 
obey ``twisted''  boundary conditions (BCs):
\beq
\psi_i (x^\mu,\cy+L) = T\left[\psi_i(x^\mu,\cy)\right] \equiv e^{i\alpha_i} \psi_i(x^\mu,\cy)\,,
\label{bc}
\eeq
where $x^\mu,~\mu=0,\ldots,3$ denote the coordinates of the 4D
Minkowski space-time (${\cal M}_4$) and $T$ is the twist operator. 
We will also assume that the fermionic mass parameters $M_i$
are positive and choose a convention where the
Dirac matrices $\gamma^M$ in 5D are the usual ones 
for $ M \not=4$ while $ \gamma_{M=4} = i\gamma_5$; we will also use the 
metric diag$(1,-1,-1,-1,-1)$.

The action is invariant under the local U(1) transformation
\beq
\psi_i(x,y) \to e^{-i\e5 q_i\Lambda(x,y)}\psi_i(x,y),\lsp
A_M(x,y) \to A_M(x,y)+\partial_M\Lambda(x,y)\,.
\label{u1symm}
\eeq
In addition, the Lagrangian is symmetric under the 
5D CP transformations~\cite{Shimizu:1984ik}
\beq
x^M\to\epsilon^M x^M, \quad 
A^M \to - \epsilon^M A^M ,\quad 
\psi_i \to \eta_i \gamma^0 \gamma^2 \psi_i^{\star},\quad
|\eta_i|=1\,,
\label{gaugcp}
\eeq
where $ \epsilon^{0,4} = - \epsilon^{1,2,3} = +1$ and there is no summation over $M$.

It is  straightforward to expand the fields in Fourier series,
leading to an infinite tower of fields propagating in ${\cal M}_4$,
\beq
\psi_i(x,y) = \frac{1}{\sqrt{L}}\sum_{n=-\infty}^\infty \psi_{i,\, n}(x)\, e^{i \bar\omega_{i,\, n} \cy},
\quad
A^M(x,y)  = \frac{1}{\sqrt{L}}\left[\sum_{n=-\infty}^\infty  A^M_n(x)\, e^{i \omega_n \cy} + a \delta^M_4\right]\,,
\eeq 
where $\omega_n = 2\pi n/\rad$ and $ \bar\omega_{i,\, n} = \omega_n + \alpha_i/\rad$.
The fields associated with the $M=4$ component 
of the gauge field become 4D scalars,
which raises the interesting possibility that 
$A_{M=4}$ may acquire a non-zero \vev; this, in fact,
is known to occur~\cite{Hosotani:1983xw}. In this case (\ref{gaugcp}) 
suggests that this is also a sign of spontaneous
CPV~\footnote{An attempt to generate CPV in a similar spirit has also been
considered in Refs.~\cite{Cosme:2002zv}.}, an expectation that is indeed confirmed, as we
will see in the following section.

It should also be emphasized that the BCs
(\ref{bc}) are not symmetric under CP
(unless $\alpha_i=0,\pm \pi$),
and this is an {\it additional}  source of explicit\footnote{In order to see that
the BCs  generate explicit CPV, it is sufficient to reformulate the theory
in terms of the periodic field $\psi^{'}(y)\equiv e^{-i \alpha y/L}\psi(y)$. Then
the BCs preserve CP but CP-violating interactions appear explicitly in the 5D Lagrangian.}  CPV, present
even if $\avev=0$. Note that the twist operator T~\cite{Scherk:1978ta} does not commute
with CP (which is a symmetry of (\ref{lagqed})), therefore the CP violation
by the boundary terms is an example of the Scherk--Schwarz breaking mechanism~\cite{Scherk:1978ta}.

Let us first focus on the fermionic piece of the Lagrangian (\ref{lagqed}). 
Integrating over the $\cy$ coordinate we find
\beq
\lcal_\psi = \sum_{i n} \bar \psi_{i,\, n} \left[ i \gamma^\mu\partial_\mu - M_i 
     +i \gamma_5 \mu_{i,\, n} \right] \psi_{i,\, n} 
   -e \sum_{i, l, n} q_i 
 \bar \psi_{i,\, l}  \left( \not\!A_{l-n} + i A^4_{l-n}\gamma_5 \right) \psi_{i,\, n} \,,
\eeq
where $\mu_{i,\, n} \equiv \left[2 \pi n + (\alpha_i + e q_i L a)\right]/L $, with 
$e\equiv \e5/\sqrt{L}$ the 4D gauge coupling.
In order to diagonalize the fermion mass term
we define the angles $ \theta_{i,\, n}$ by
\beq
\tan( 2 \theta_{i,\, n}) = { \mu_{i,\, n} \over M_i}; \quad
| \theta_{i,\, n}| \le \pi/4
\label{theta}
\eeq
and replace\footnote{
The chiral rotation of the fermions induces an 
$\epsilon_{\mu\nu\sigma\rho}F_{\mu\nu}F^{\sigma\rho}$ 
term in the Lagrangian;
however, in the Abelian case considered here, this is a total derivative and it can be dropped.} 
$ \psi_{i,\, n} \to \exp( i \gamma_5 \theta_{i,\, n}) \psi_{i,\, n} $.
From this we find that the physical fermion masses are
$ m_{i,\, n} = \sqrt{ M_i^2 + \mu_{i,\, n}^2} $, 
while the interactions with the gauge fields read
\bea
\lcal_{A\psi} &=&
-e\sum_i q_i \Big\{A_\mu\sum_k \bar\psi_{i,\, k} \gamma^\mu \psi_{i,\, k} 
+\sum_{k\neq l} A_{\mu\; k-l}\bar\psi_{i,\, k} \Gamma_{i,\, kl}^{(v)}\gamma^\mu\psi_{i,\, l}
\Big\}\,,\label{vect}\\
\lcal_{\varphi\psi} &=&
-e \sum_i q_i \Big\{\varphi \sum_k \bar\psi_{i,\, k} \Gamma_{i,\, k}^{(\varphi)} \psi_{i,\, k} 
+ \sum_{k\neq l} A_{4\; k-l}\bar\psi_{i,\, k} \Gamma_{i,\, kl}^{(s)}\psi_{i,\, l} 
\Big\}\,,
\label{scal} 
\eea
where $\varphi \equiv A_{4\; 0}$, $A_\mu \equiv A_{\mu\ 0}$ and
\beq 
\Gamma_{i,\, k}^{(\varphi)} \equiv -i \gamma_5 e^{2i\gamma_5 \theta_{i,\, k}}, \lsp
\Gamma_{i,\, kl}^{(s)} \equiv -i\gamma_5 e^{i\gamma_5 (\theta_{i,\, k}+\theta_{i,\, l})}, \lsp
\Gamma_{i,\, kl}^{(v)} \equiv  e^{i\gamma_5 (\theta_{i,\, k}-\theta_{i,\, l})} \,.
\eeq

It is evident that $A_\mu$ corresponds to the 4D photon. However, the 
field $\varphi$ is a new, physical, low-energy 
degree of freedom whose Yukawa couplings appear to be CP-violating.
As we will show shortly, this naive conclusion is 
 incorrect in general. The couplings of $A_{4\; n}$ and $A_{\mu\; n}$  also  appear
to violate CP.

The 5D gauge transformation (\ref{u1symm})  in terms of KK modes 
implies $ A^4_k \to A^4_k + i \omega_k \Lambda_k$, where
$\Lambda(x,y)=L^{-1/2}\sum_{n=-\infty}^{+\infty} \Lambda_n(x) e^{i\omega_n y}$,
which shows that,
while $A^4_{k\neq 0}$ can be removed by an appropriate gauge choice,
$\varphi=A_{4\; 0}$ is a gauge singlet\footnote{
This is a consequence of the compactification of the $x^4$ direction;
in an uncompactified space one could always choose the $A^4(x,y)=0$
gauge. Note also that in the case of compactification on the 
orbifold $S^1/Z_2$, $\varphi$ disappears as
a consequence of the requirement of the antisymmetry of $A_4$ under $Z_2$: $y\to-y$.
Therefore CP cannot be violated spontaneously; however, if BCs are not symmetric under CP,
$A_{\mu\; n}$ and $A_{4\; n}$ would still have CP-violating couplings with fermions.}. 
Because of this, even though  the $ \varphi $ mass 
$ m_\varphi$ vanishes at tree level (see (\ref{kin}) below), 
it will receive calculable finite corrections at higher orders 
in perturbation theory.

It is worth noting that even if  $A_4(x,y)=\varphi(x)$ by a choice
of gauge, there still remains a residual $y$-dependent discrete 
gauge freedom
\beq
A_4\to A_4 + \frac{2 \pi n_i}{e_5q_i L}\,, \lsp \psi_i \to e^{-i \frac{2 \pi n_i}{L}y}\psi_i\,,
\quad n_i = 0 , \pm1 , \cdots,
\label{res}
\eeq
{\em provided} $q_1/q_2$ is a rational number.
In this case there exist some discrete constant values of $\varphi $
($=2\pi n_i/(\e5 q_iL) $) that can be removed completely. 
Note also that $\alpha_i + \e5 q_i A_4 $ is invariant under 
(\ref{res}); this will be relevant when we discuss 
the one-loop effective potential for $\langle\varphi\rangle$.
 
The  physical content of KK excitations for $A_4$ can be easily revealed by adopting
the following generalization of the 4D R$_\xi$ gauge~\cite{Muck:2001yv}:
\beq
\lcal_{gf}=-\frac{1}{2\xi}\left(\partial^\mu A_\mu-\xi\partial_y A_4\right)^2\,.
\eeq
Decomposing into KK modes, one can write the kinetic part
of  the Lagrangian density in  the following form
\bea
\lcal_A + \lcal_{gf}&=&\frac12\sum_n\left\{ A^\mu_n\left[(\square + \omega_n^2) g_{\mu\nu}
- (1-\xi^{-1})\partial_\mu\partial_\nu\right] A^\nu_{-n}
- A^4_n(\square + \xi \omega_n^2)A^4_{-n}\right\}\,.
\label{kin}
\eea
It is then clear that $A^4_n$ ($n \neq 0$) are the would-be Goldstone bosons that 
become a longitudinal component of $A^\mu_n$, while $\varphi=A_{4\;0}$ is 
a physical massless
scalar.
Note  that even for $ \avev \not=0 $, the 4D $\ui$ gauge symmetry
remains unbroken, so that the 4D photon $ A_{\mu\ 0} $ remains massless.

%33333333333333333333333333333333333333333333333333333333333333333333333333333333333333333333
\section{The effective potential}
\label{efpot}
The above discussion raises the possibility that $\varphi$ will 
acquire a non-vanishing \vev\ $a\equiv \langle \varphi\rangle$. In order to determine the
conditions under which this occurs, we evaluate
the corresponding effective potential to one loop. We will adopt
dimensional regularization for the $d^4p$ integral together
with a summation over the infinite tower of KK modes. 

After dropping an irrelevant constant contribution,
and using dimensional regularization for the $d^4p$ integral, we find
\beq
V(M;\omega)=\frac{1}{32\pi^6L^4}\left[x^2Li_3(re^{-x})+3xLi_4(re^{-x})+3Li_5(re^{-x}) + {\rm H.c.}\right]\,,
\label{vfer}
\eeq
where $x\equiv L M$, $\omega=(\alpha + e\qp L a)/L$, $r=\exp(iL \omega)$, and $Li_n(x)$ is 
the standard polylogarithm function.
Note that, as a consequence of the hermiticity, the potential is a symmetric function of $\omega$:
$V(M;\omega)=V(M;-\omega)$.
We will consider a theory that contains two fermionic fields, so the 
total effective potential reads
\beq
V_{eff}(a)=\sum_{i=1,2}V(M_i;\omega_i)\,.
\eeq

The total effective potential is not periodic in $a$ unless
$q_1/q_2$ is a rational number $n_1/ n_2$, in which
case the period is  
$T=2\pi n_1/(eq_1L)=2\pi n_2/(eq_2L)$. 
This property is a consequence of the residual gauge invariance (\ref{res})
present when $ q_1/q_2$ is rational.

It is worth discussing what would happen if we had just one fermionic field $\psi$.
In this case the minimum of $V$ is at $L\omega = \pi (2l+1)$ for integer $l$, but since 
$\alpha$ is defined modulo $2\pi$ we can choose the minimum $\alpha + e\qp L a =\pi$. 
We can also eliminate $ \alpha $ from (\ref{bc}) by the following field 
redefinition:
\beq
 \psi^\prime (x,y) = e^{  - i \alpha y/L}\psi (x,y),
\qquad
\e5 \qp A_M^\prime (x,y)=  \e5 \qp A_M (x,y) + \alpha \delta_{M,4}\,,
\eeq
so that $ \psi'$ and $A'$ are periodic in $y$ with period
$L$.
We then expand around the 
vacuum $\e5 \qp \langle A_M' \rangle=(\pi/L)\delta_{M4}$ by shifting
the gauge field $\e5 \qp A_M' (x,y) \to \e5 \qp A_M' (x,y) +( \pi/L)\delta_{M4}$,
and again redefine the fermion fields, so that the effect of this shift disappears 
from the Lagrangian density:
\beq
\chi (x,y) = e^{i \pi y/R} \psi^\prime (x,y)\,,
\eeq
which is antiperiodic in $y$, $ \chi (x,y+L) = - \chi (x,y) $.

Through this series of field redefinitions, we have 
shown that the original theory is equivalent to one where
the gauge field has a vanishing \vev\ (hence, no spontaneous CPV)
and also the fermionic field has CP-invariant BCs; consequently the theory predicts 
no CP-violating effects. 
However, in Eqs.~(\ref{vect}) and (\ref{scal}) we have noted the presence of CP-violating 
couplings of $\varphi$, $A_{\mu\; n}$ and $A_{4\; n} $ even if only one fermion is present;
this therefore deserves further explanation. 

At the minimum of $V$, $ \mu_n  = \pi(2n+1)/L$ and 
$m_n=m_{-n-1}$, so that any unitary transformation $U$ acting 
on the $(\psi_n,\psi_{-n-1})$ subspace will leave the 
corresponding kinetic terms invariant.
This allows for a generalized definition of the CP transformation:
\beq
\psi_i \stackrel{{\rm CP}}{\lra} U_{ij} C \overline{(\gamma_0\psi_j)}^T\,, i,j=n,-n-1\,,
\label{newcp}
\eeq
where $C\gamma_\mu C^{-1}=-\gamma_\mu^T$.
Choosing $ U = \sigma^1 $ (the usual Pauli matrix)
one can easily see that in fact, the couplings of
$\psi_n,\psi_{-n-1}$ are invariant under CP as defined in 
(\ref{newcp}).

\begin{figure}[h]
\centering
\includegraphics[width=12cm]{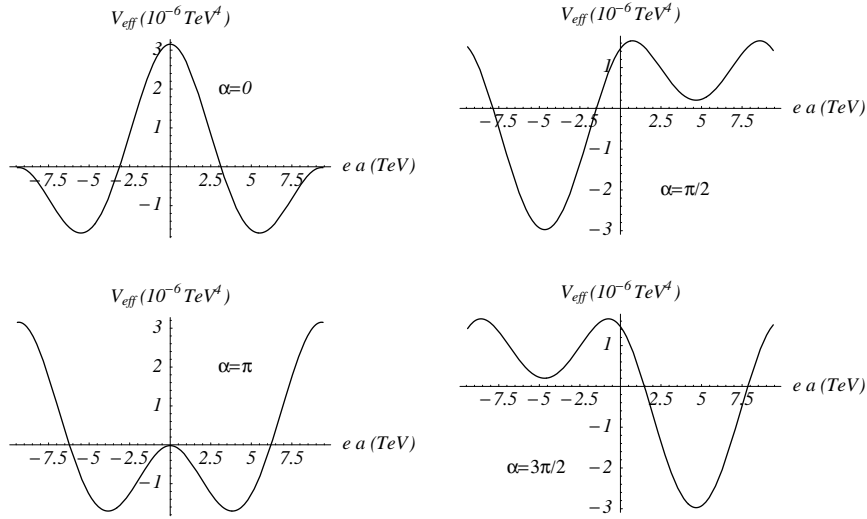}
\caption{The effective potential $V_{eff}$ in units of $10^{-6} ~\tev^4$
for $L^{-1}=0.3 ~\tev$, $M_1=0.2 ~\tev$, $M_2=0.005 ~\tev$, $q_1=2/3$, $q_2=-1/3$ 
and four choices of the twist angle $\alpha=0,\pi/2,\pi,3\pi/2$ (staring from the upper left plot
and moving clockwise) is plotted as a function of $e \, a$
in units of $\tev$.}
\label{effpot}
\end{figure}

The situation can be different if a second fermion is present. Then, following the steps
described above for the case of a single fermion, it can be shown that
without losing any generality we can adopt the 
BCs $ \psi_1(y+L)=\psi_1(y) $,
$ \psi_2(y+L)=\exp(i\alpha)\psi_2(y)$. 
The condition for an extremum is 
\beq
\frac{\partial V_{eff}}{\partial a}=e\sum_{i=1,2}q_i
\frac{\partial V(M_i;\omega_i)}{\partial \omega_i}=0
\label{extre}
\eeq
and leads to a CP-conserving vacuum  when the minimum of $V_{eff}$ is at
$\omega_i=0,\pi/L$. In general, however, the minimum is located elsewhere,
opening a possibility for spontaneous CPV. As we have seen,
at least two fermions are necessary to observe CPV in 5D QED compactified on a circle.
In this case the KK modes of both fermions will have CP-violating Yukawa couplings to $\varphi$,
$A_{\mu\; n}$ and $A_{4\; n}$.
In Fig.~\ref{effpot} we plot the effective potential for various choices of the twist angle
$\alpha$ as a function of $e\, a$. Note that $V_{eff}$ 
is a symmetric function of $e\,a$ when $ \alpha =0,\pi $ 
(technically this is a consequence of the symmetry of $V^{(L)}(M_i;\omega)$ under $\omega \to - \omega$).
This is the case of CP-symmetric BCs (\ref{bc})~\footnote{While it is trivial for $\alpha=0$, 
one can, adopting similar arguments as above, show that also for $\alpha=\pi$ there are no 
CP-non-invariant interactions in the effective theory.}, 
since the Lagrangian is invariant under CP;
therefore the whole theory is CP-symmetric. Under CP, $a \to -a$, and therefore the observed symmetry
of the effective potential is precisely a consequence of CP invariance.
Consequently, choosing any of the two degenerate vacuum leads to spontaneous CP violation.
For other (i.e. CP-asymmetric) choices of $\alpha$, the effective potential is not invariant
with respect to $a\to -a$; therefore, even though $a\neq 0$ at the minimum, CP is explicitly violated 
in those cases.

In the case of two fermionic fields there are two observable CP-violating parameters,
 $\alpha$ and $\langle A_4 \rangle$. In general, for $N_f$ fermions
there will be  $N_f$ CP-violating parameters:
$N_f-1$ twist angles and $\langle A_4 \rangle$. If, for instance, all the fermions 
are periodic in $y$ ($\alpha_i=0$), only 
 $\langle A_4 \rangle$ will parametrize all CP-violating effects. 

%444444444444444444444444444444444444444444444444444444444444444444444444444444444444444444444
\section{Phenomenology}

The most striking consequence of CPV in our model will be a prediction for  
a non-zero fermionic electric dipole moment (EDM) $d$ 
defined through the following effective $\gamma \bar{\psi}\psi$
vertex $ \langle p' | j^\mu_{EM} | p \rangle = -(d/e)
\bar{u}(p')\sigma^{\mu\nu}\gamma_5(p' - p)_\nu u(p)$ 
where $p,~p'$ are on shell and the limit $ p' \to p $ is assumed.
In our model a  non-zero EDM is generated  already at the one-loop
level (in the SM at least three loops are required).
For the fermion $\psi_i$, the  diagram involving $ \varphi $ yields\footnote{An analogous contribution 
appears within the Two Higgs Doublet Model (2HDM), see 
Refs.~\cite{Barger:1996jc,Atwood:2000tu}.}
\beq
d_{i\;0}=  - {(e q_i)^3 c\up+_{i,\;0} \over 16 \pi^2 m_{i,\; 0}}
 J\up s(m_\varphi^2/m_{i,\;0}^2,1)\,,
\label{s_edm}
\eeq
while the contributions from the $n$-th modes circulating in the loop equal
\bea
d^{(v)}_{i,\; n} &=&  \frac{(e q_i)^3 c\up-_{i,\;n}}{4 \pi^2} 
\frac{ m_{i,\;n}} {m^2_{i,\; 0}} J\up v(x_{i,\;n},y_{i\;n})\quad
{\rm ~from~} A^\mu_n {\rm ~exchange} \,,\cr
d^{(s)}_{i,\; 0} &=& - \frac{(e q_i)^3 c\up+_{i,\;n}}{16\pi^2} 
\frac{ m_{i,\;n}} {m^2_{i,\; 0}} J\up s (x_{i,\;n},y_{i\;n})\quad
{\rm ~from~} A^4_n {\rm ~exchange}\,,
\eea
where $ c_{i,\;n}\up\pm = \pm M_i(\mu_{i,\, n} \pm \mu_{i,\, 0}) / (m_{i,\;n} m_{i,\;0})$,
$ x_{i\;n} = (\omega_n/m_{i,\;0})^2, ~ y_{i\;n} = (m_{i,\;n}/m_{i,\;0}) ^2 $,
and
\bea
J\up s(x,y) &=&
1 + {x - y + 1\over 2} \ln \left(y\over x\right) + \left({2 x\over \rho} - \rho\right) \Theta \,,\cr
J\up v(x,y) &=&
-1 + {y - x\over 2} \ln \left({y\over x}\right) + (\rho - \cot\Theta)\Theta\,,
\eea
with $ \rho^2 \equiv 4 xy -(x+y-1)^2 $ and $\tan\Theta \equiv \rho/(x+y-1)$.

The total EDM of the $i$-th zero-mode fermion is then
\beq
d_i=d_{i,\, 0} + {(e q_i)^3 \over 16 \pi^2 m_{i,\, 0}^2} 
\sum_{n\neq 0} m_{i,\, n} \left[4 c\up-_{i,\;n} J\up v (x_{i,\;n},y_{i\;n}) - 
c\up+_{i,\;n}  J\up s(x_{i,\;n},y_{i\;n})\right]\,.
\label{edm}
\eeq
Note that for large $n$, $ c\up\pm J^{(s/v)}(x_{i,\;n},y_{i\;n}) \sim  1/n+{\cal O}(1/n^2)$,
so that $d_i$ will be finite after symmetric summation over $n$. It follows that
the EDM is finite and therefore insensitive to the cut off of the 5D theory.

\begin{figure}[h]
%\centering
\includegraphics[width=7.8cm]{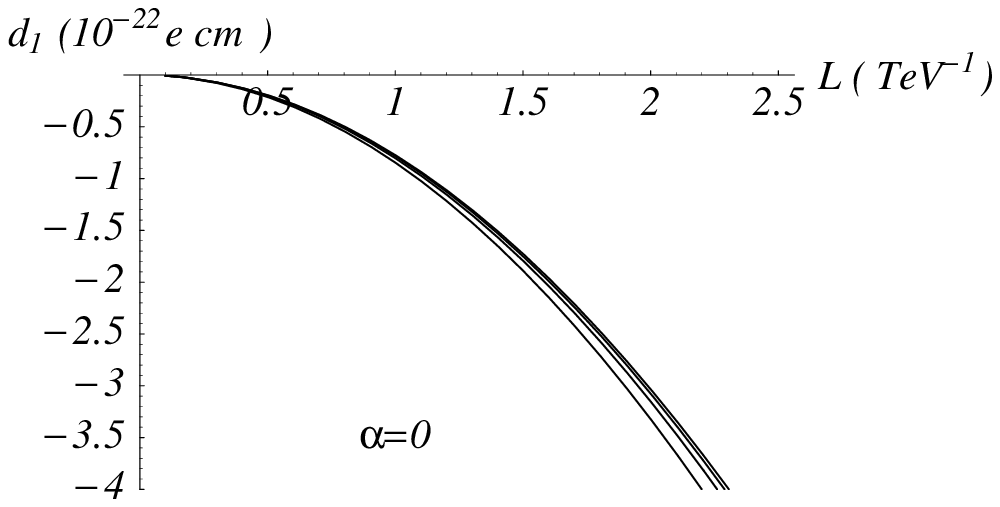} \quad
\includegraphics[width=7.8cm]{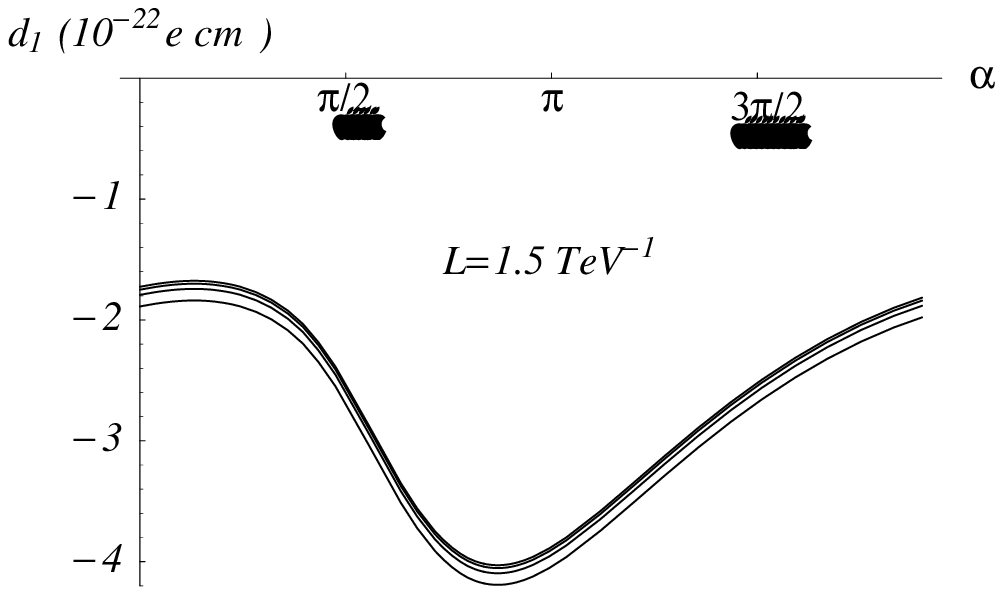}
\caption{Left: The fermionic EDM, $ d_1$, in units of $10^{-22}$ 
e cm for the zero-mode of the fermion $i=1$
as a function of the compactification length $L$ in units of $\tev^{-1}$ 
for $\alpha=0$. Right: Same, as a function of $ \alpha$ for $ L=1.5~\tev^{-1}$.
The curves from the bottom to the top
correspond to the number of modes included in (\ref{edm}) varying from $|n|=1$ to $|n|=4$;
the fast convergence of the series is evident. Note that the mass 
of the zero-mode also varies with $L$. We used 
$e=\sqrt{4\pi\alpha_{QED}}$ and the same parameters as in Fig.~\ref{effpot}.}
\label{edmplot}
\end{figure}

In Fig.~\ref{edmplot} we plot the fermionic EDM as a function of the compactification scale $L$.
The parameters have been adjusted in such a way that the model has the  mass scales and the coupling constants
of the same order as those that are present in the SM.
We have chosen for illustration to plot the EDM of the zero-mode of the  $i=1$ fermion.
Note  that the mass of the zero-mode depends on $L$; 
for parameters adopted here (with $\alpha=0$, i.e. for the case of spontaneous CP), 
it varies from $\sim 37~\tev$ for $ L=0.1~\tev^{-1}$, to $\sim 1.5~\tev$ 
when $L=2.5~\tev^{-1}$; for these values, $ m_\varphi$  ranges from $\sim 88~\gev$ 
to $\sim 3.5~\gev$.
The leading contribution to $d_i$ comes from the $\varphi$ exchange; the contribution of the non-zero
modes is of opposite sign and smaller by a factor ${\cal O}(5)$.

In Fig.~\ref{edmplot} the positive \vev\ of $A_4$ was chosen (see Fig.~\ref{effpot}) for $\alpha=0$. 
It is worth noticing that EDM, as a CP-odd quantity, would be of the opposite sign if the other (negative)
\vev\ was chosen. 

%5555555555555555555555555555555555555555555555555555555555555555555555555555555555555
\section{Conclusions}

We have shown that QED in $(1+4)$-dimensional space-time, with the fifth 
dimension compactified on a circle, leads to CP violation.
Depending on fermionic boundary conditions, CPV may be either explicit, or
spontaneous via the Hosotani mechanism. The new possibility
of CP breaking by fermionic, twisted boundary conditions has been emphasized
and demonstrated explicitly by derivation of CP-violating effective couplings.
The fifth component of the gauge field acquires  (at the one-loop level) 
a non-zero vacuum expectation value. We have shown that in the presence of two fermionic fields, this
leads to spontaneous CPV in the case of CP-symmetric boundary conditions.
The one-loop effective potential for $A_{4\; 0} $ has been calculated and its features  
have been discussed in the presence of two fermionic fields. 

The most striking feature of the model considered here 
is the presence of the light scalar $\varphi$, which
has CP-violating Yukawa couplings similar
to those present in the scalar sector of the 2HDM model.
The presence of CP-violating couplings leads to 
a non-zero EDM, which was calculated at the one-loop level for a zero-mode fermion. 
This effect can be used to test the mechanism for CPV present in our model. 

There are several other
observables, originally developed to investigate extended Higgs sectors,
which can also be used to detect the presence of a light scalar 
(regardless of whether its couplings conserve or violate CP) such as $ \varphi$.
For example, aside from the fermionic
electric and magnetic dipole moments, one also has
$\Gamma[\Upsilon \to \varphi \gamma]$ and $BR[b\to \varphi s]$. 
Experimental constraints on all such quantities would impose some restrictions
on the parameters of the model. We will present the results of such an investigation
in a separate publication, where we will consider a
more realistic non-Abelian theory.

\vspace*{0.5cm}
% AAAAAAAAAAAAAAAAAAAAAAAAAAAAAAAAAAAAAAAAAAAAAAAAAAAAAAAAAAAAAAAAAAAAAAAAAAAAAAAAAAAAAAAAAAAA
\centerline{\bf ACKNOWLEDGEMENTS}
\vspace*{0.5cm}
This work is supported in part by the State
Committee for Scientific Research (Poland) under grant 5~P03B~121~20,
and by funds provided by the U.S. Department of Energy under grant No.
DE-FG03-94ER40837.

\end{document}